\documentclass[twocolumn]{article}
\usepackage{bagrow}
\usepackage{fullpage}
\usepackage{amsmath}
\usepackage{amssymb}
\usepackage{amsthm}
\usepackage{graphicx}
\usepackage[]{newtxtext}
\usepackage[bigdelims]{newtxmath}
\usepackage{booktabs}
\usepackage[usenames,dvipsnames,svgnames,table]{xcolor}
\usepackage{xcolor}
\usepackage[normalem]{ulem}
\usepackage[final]{pdfpages}

\usepackage{mathtools}
\usepackage{subfigure}
\usepackage{soul}
\usepackage{multirow}
\usepackage{enumitem}
\usepackage[usenames,dvipsnames,svgnames,table]{xcolor}

\newcommand{\E}{\operatorname{E}} 
\newcommand{\avg}[1]{\E\left[#1\right]}
\DeclarePairedDelimiter\abs{\lvert}{\rvert}%

\makeatletter
\let\oldabs\abs
\def\abs{\@ifstar{\oldabs}{\oldabs*}}
\makeatother

\definecolor{MediumGreen}{HTML}{008800}

\author[1,2]{Abigail Hotaling}
\author[1,2,*]{James P.~Bagrow}
\affil[1]{Department of Mathematics \& Statistics, University of Vermont, Burlington, VT, United States}
\affil[2]{Vermont Complex Systems Center, University of Vermont, Burlington, VT, United States}
\affil[*]{\corrauthinfo{james.bagrow@uvm.edu}{bagrow.com}}

\date{\today}%

\AtBeginEnvironment{abstract}{\singlespacing}
\title{Efficient crowdsourcing of crowd-generated microtasks}

\begin{document}

\twocolumn[
  \begin{@twocolumnfalse}
    \maketitle
    \begin{abstract}
Allowing members of the crowd to propose novel microtasks for one another is an effective way to combine the efficiencies of traditional microtask work with the inventiveness and hypothesis generation potential of human workers.
However, microtask proposal leads to a growing set of tasks that may overwhelm limited crowdsourcer resources.
Crowdsourcers can employ methods to utilize their resources efficiently, but
algorithmic approaches to efficient crowdsourcing generally require a fixed task set of known size.
In this paper,
we introduce \emph{cost forecasting} as a means for a crowdsourcer to use efficient crowdsourcing algorithms with a growing set of microtasks.
Cost forecasting allows the crowdsourcer to decide between eliciting new tasks from the crowd or receiving responses to existing tasks based on whether or not new tasks will cost less to complete than existing tasks, efficiently balancing resources as crowdsourcing occurs. 
Experiments with real and synthetic crowdsourcing data show that cost forecasting leads to improved accuracy.
Accuracy and efficiency gains for crowd-generated microtasks hold the promise to further leverage the creativity and wisdom of the crowd, with applications such as generating more informative and diverse training data for machine learning applications and improving the performance of user-generated content and question-answering platforms.
    \end{abstract}
\bigskip\bigskip

  \end{@twocolumnfalse}
]

\section{Introduction}

Crowdsourcing platforms enable large groups of individual crowd members to collectively provide a crowdsourcer with new information for many problems~\cite{brabham2008crowdsourcing,kittur2013future} such as completing user surveys~\cite{behrend2011viability}, generating training data for machine learning models~\cite{snow2008cheap,wagy2017crowdsourcing}, or powering citizen science programs~\cite{kamar2012combining,franzoni2014crowd}.
The work performed by the crowd is often used by researchers and firms to address problems that remain computationally challenging.
Yet incorporating humans into a problem domain introduces new challenges: workers must be paid and even volunteers should be properly incentivized, bad actors or unreliable crowd members should be identified, and care must be taken to efficiently and accurately aggregate the response of the crowd.
Algorithmic crowdsourcing focuses on computational approaches to these challenges, allowing crowdsourcers to maximize the accuracy of the data generated by the crowd while also efficiently managing the costs of employing the crowd.

Despite the potential challenges, engaging a crowd is often invaluable, as
crowd participants are capable of creative ideation in a way that computational methods are not, and they can generate novel ideas or new tasks beyond those designed by the crowdsourcer.
\emph{Crowd-generated microtasks} are an important avenue for this creativity to manifest~\cite{mcandrew2017reply,liu2018incentivizing}:
The members of the crowd may be asked to not simply provide responses to given tasks, but also to propose new tasks to give to other crowd members.
Combining  task proposal with microtask work provides the crowd a simple vehicle to introduce their own new ideas and hypotheses~\cite{mcandrew2017reply}, while still leveraging the known efficiency of microtask work~\cite{kittur2011crowdforge}.

Crowd-generated microtasks have been used for a number of practical crowdsourcing applications.
Examples include feature generation for machine learning methods~\cite{bongard2013crowdsourcing,wagy2017crowdsourcing}, used to explore novel predictors of childhood obesity and home energy use;
the `verify' step of Soylent's Find-Fix-Verify algorithm~\cite{bernstein2010soylent}, enabling prose writing to leverage microtask work within a crowd-powered word processor;
crowdsourced creation of knowledge networks~\cite{mcandrew2017reply,berenberg2018efficient}, allowing for an improved understanding of causal attribution; and contributing new questions to a growing user survey~\cite{salganik2015wiki},
used to generate and vote upon novel ideas for New York City's government to improve the welfare of its citizens.
In all these examples, as new content are generated, the crowdsourcer is left to manage a growing set of simple, associated microtasks such as answering multiple-choice questions or voting for proposed ideas.

Another popular application of crowd-generated microtasks is question-answering (QA) websites, online communities where members can pose new questions they wish answered and provide answers to questions posed by other members~\cite{Zhang:2007:ENO:1242572.1242603,Bian:2008:FRF:1367497.1367561}.
Although asking and answering questions are often open-ended tasks, microtasks are a key component of administrating a QA platform, with the platform provider instantiating any number of additional microtasks for purposes such as labeling or classifying content~\cite{liu2018incentivizing}.
One example of such a microtask, which is often embedded as part of other tasks, is a yes/no survey showing members a question paired with a user-submitted answer and asking if this question is now sufficiently answered.

Allowing the crowd to generate tasks can lead to a growing set of tasks and this growth, even slow growth, can eventually overwhelm the crowdsourcer's resources and the majority of tasks will remain unseen by the crowd~\cite{mcandrew2017reply}.
Thus a crowdsourcer using crowd-generated microtasks must use resources efficiently.
Algorithmic crowdsourcing addresses efficiency with methods for the crowdsourcer to allocate tasks to the crowd and efficiently and accurately infer answers for given microtasks~\cite{dawid1979maximum,chen2013optimistic}.
However, most allocation algorithms assume a fixed set of tasks 
to distribute to the crowd.
Our goal here is to study how algorithmic crowdsourcing methods can best be used for crowdsourcing problems with crowd-generated microtasks. 
We introduce a decision process---\emph{cost forecasting}---that enables a crowdsourcer to decide online whether to grow the set of tasks or receive responses to existing tasks. 
For problems where the crowdsourcer can make this choice, this provides a means to apply efficient algorithms to crowd-generated microtasks, allowing the crowdsourcer to achieve high quality work on tasks even when the set of tasks is open-ended.

The rest of this paper is organized as follows.
In Sec.~\ref{sec:background} we provide background describing the crowdsourcing problem model we consider, existing methods for crowdsourcing crowd-generated microtasks, and prior work on efficient crowdsourcing (budget allocation) algorithms.
We introduce cost forecasting in Sec.~\ref{sec:costforecasting} and derive probabilistic estimators using it on our problem model. 
We report in Sec.~\ref{sec:methods} (Materials and Methods) and Sec.~\ref{sec:results} (Experiments) our results using real and synthetic crowdsourcing data to investigate the accuracy of collected data when crowdsourcing with cost forecasting. 
We also describe in Sec.~\ref{sec:methods} how to simulate crowd-generated microtasks using pre-existing crowdsourced datasets.
Lastly, we conclude in Sec.~\ref{sec:discussion} with a discussion of this work and its applications, including the limitations of our study and promising directions for future research.

\section{Background}
\label{sec:background}

Here we describe the problem model we employ in our study to represent crowdsourcing tasks, describe prior research on crowd-generated microtask crowdsourcing, as well as provide details on existing methods for crowdsourcing microtask data under budget constraints.
\subsection{Problem model and existing work}
\label{subsec:background:problemmodel}

We focus on problems where crowd members propose binary labeling tasks as a representative model for individual microtasks, as is standard practice in algorithmic crowdsourcing.
In the context of crowd-generated microtasks, workers can introduce novel microtasks for other workers to label, leading, perhaps after appropriate validation, to a growing set of labeling tasks.
For example, when crowdsourcing causal attributions~\cite{berenberg2018efficient}, a worker may introduce a novel microtask by posing a new question (\emph{Do you think that viruses cause sickness?}) which then becomes a new yes/no binary labeling microtask for other crowd workers.
While binary labeling is a simplification of the nuance of many real-world crowdsourcing tasks, binary labeling can represent image categorization tasks or even basic survey questions, and can be readily generalized to categorical labeling tasks such as multiple choice questions, although those tasks can also be binarized (see \cite{li2016crowdsourcing}).

Let $z_i \in \{0,1\}$ be the true but unknown label for task $i$ and let $y_{ij}$ be the response provided by worker $j$ when given task $i$.
We define the associated task parameter $\theta_i \equiv \mathrm{Pr}(z_i = 1)$ as the unknown probability that the true label for task $i$ is 1.
Multiple workers are typically asked to respond to a given task, allowing us to aggregate their responses for improved accuracy; we assume that workers respond independently so that the $\{y_{ij}\}$ are iid for a given $i$. 
To track the response tallies for task $i$, let $a_i$ and $b_i$ be the total number of `$+1$' and `$0$' responses, respectively, for $i$, and let $n_i = a_i + b_i$ be the total number of responses received for $i$.
As responses are gathered, these tallies will change, so $a_i$, $b_i$, and $n_i$ are considered functions of time $t$, where we track `time' as the number of responses received across all workers and tasks ($t = \sum_i n_i(t)$). 
We can estimate $\theta$ with $\hat{\theta} = {a_i}/{n_i}$.
The final goal is to infer the true label of the task accurately, i.e., develop $\hat{z_i} \approx z_i$ using the responses $\{y_{ij}\}$ for task $i$.

Most work on efficient crowdsourcing assumes a fixed set of tasks but some studies have considered task growth.
The work of Sheng, Provost \& Ipeirotos \cite{sheng2008get} considers the idea of soliciting new training examples (labeling tasks) from the crowd, and discusses strategies for how often to request new tasks depending on the cost of receiving a new task relative to the cost of receiving a response to an existing task.
However, the focus on their work is on how many responses a single task requires, as multiple responses are typically used to overcome noisy workers, and they do not consider the cost to complete a task (something we will focus on; Sec.~\ref{sec:costforecasting}), only the cost on a per-response basis.
Likewise, the recent work of Liu and Ho~\cite{liu2018incentivizing} studies task growth using a multi-armed bandit approach, where the arms of the bandit increase over time.
They assume the crowdsourcer is not able to control when new tasks are generated, however, and neither study considers the use of efficient allocation methods for guiding workers to tasks when costs are constrained by a budget.
Of course, returning to the example of a QA platform, users typically submit questions on their own, but any QA site can implement an approval process allowing the site to control the rate of new questions. %
To the best of our knowledge, crowdsourcing a growing set of tasks when efficient allocation methods are used to complete those tasks has not been studied.

\subsection{Efficient allocation methods}
\label{subsec:background:efficientallocation}

Often a crowdsourcer must accurately infer the $z_i$ labels under budget constraints, as only finite resources (such as time or money) will be available to support the crowd.
For simplicity, we assume a crowdsourcer has a total budget of $B$ requests that can be elicited from the crowd.
The budget then imposes the constraint $\sum_i n_i(t) \leq B$ for all $t \leq B$.
This constraint becomes especially challenging for a growing set of tasks, since the finite budget must be spread out over an increasing number of individual tasks.

Crowdsourcing allocation methods~\cite{chen2013optimistic,karger2014budget,li2016crowdsourcing} have been developed to efficiently and accurately infer labels for tasks under a finite budget. 
These methods choose which tasks to give to workers with a goal of maximizing the efficiency and accuracy of the task labels the crowdsourcer will infer from the worker responses.
In this work, we apply the Optimistic Knowledge Gradient (Opt-KG) method~\cite{chen2013optimistic}. 
Opt-KG works to optimize accuracy by implementing a Markov Decision Process that chooses tasks with the largest expected improvement in accuracy. 
This method has shown improvement in accuracy when applied to finite budget crowdsourcings~\cite{chen2013optimistic}.
Opt-KG focuses on optimizing \emph{overall} accuracy, which makes it particularly beneficial for applying to crowd-generated microtasks and is the reason we focus on it in this work (see also our discussion of Opt-KG and other methods in Sec.~\ref{sec:discussion}).
Further, Opt-KG has no parameters that need to be tuned or chosen by the crowdsourcer.

Opt-KG and other allocation methods assume a fixed set of $N$ tasks.
The goal of our work here is to enable an efficient allocation method to support crowdsourcing problems where the crowd can provide new tasks to the crowdsourcer, leading to a set of tasks that grows over the duration of the crowdsourcing.

\section{Cost Forecasting}
\label{sec:costforecasting}

Here we introduce a method to enable efficient allocation methods such as Opt-KG to work with crowd-generated microtasks.
First, we extend the traditional binary labeling model for a fixed set of tasks to an open-ended problem where the crowdsourcer begins with a small seed of tasks that grows as the crowd generates novel tasks.
We then describe the components of cost forecasting including cost estimators for how many responses are needed to complete tasks and a decision rule (Growth Rule) based on those costs that allows the crowdsourcer to choose whether a crowd worker should work on an existing task or propose a new task.

\subsection{Model for crowd-generated microtasks}
\label{subsec:costforecasting:modelopenended}

The problem model given above (Sec.~\ref{subsec:background:problemmodel}) describes each of a fixed set of $N$ tasks.
Typically, allocation methods assume there is a fixed number of tasks that a crowdsourcer wishes to distribute to workers.
However, in this work we consider \emph{task growth} where the number of tasks grows as new tasks are generated by the crowd.
Growing tasks can represent the submission of new questions to a question-answering site, for example, while responding to a task represents a user answering an existing question or more simply flagging an existing question-answer pair as correct.

Let $N_t$ be the total number of tasks that exist at time $t$, where $N_0$ initial seed tasks are used to begin the crowdsourcing and we track time such that each timestep represents one request made by the crowdsourcer.
When a new task is desired at timestep $t$, a worker will be prompted to propose a new task, which is then added to the set of all tasks, and $N_{t+1} = N_t+1$. %
Later, other workers can submit responses to this new task so that a label for that task can be inferred. 
In this model, the cost of a new task generated by the crowd and the cost of a response is defined to be $f_t$ and $f_r$ units, respectively.
Depending on problem-specific considerations, the crowdsourcer can set $f_t = f_r$ or let the costs differ (see also \cite{sheng2008get}). 
In this work, we define cost units in number of responses, taking $f_t = f_r = 1$; we discuss $f_t \neq f_r$ in our discussion.
In practice, an approval process may also be needed to guarantee requirements for the new task such as appropriateness, novelty, or importance.
For simplicity, here we assume this process has already been implemented.

\subsection{Forecasting the cost to complete a task}
\label{subsec:costforecasting:forecastingcost}

Suppose at some time $t$ during the crowdsourcing that task $i$ has already received $n_i(t)$ independent $(0,+1)$ responses, of which $a_i(t)$ are +1 responses. 
Our current estimate of the task's associated parameter $\theta_i$ is $\hat{\theta}_i(t) = a_i(t) / n_i(t)$.
We can decide if task $i$ should be labeled $+1$ or labeled $0$ based on whether $\hat{\theta}_i > 1/2$ or $\hat{\theta}_i < 1/2$, but we want to minimize the probability of giving $i$ the wrong label.
This may require waiting until more responses to $i$ are gathered, so a conclusion can be drawn more safely, but we also want to avoid wasting additional responses on tasks that we can already label $i$ with an acceptable accuracy or on tasks that are too difficult (or too expensive) to answer accurately.
Thus, we need to incorporate our uncertainty in $\hat{\theta}$ given the collected data.

In general, for $n$ independent samples of a  Bernoulli random variable, the probability that our estimate $\hat{\theta}$ differs from the true value $\theta$ by at least $\epsilon$ is bounded by \emph{Hoeffding's Inequality}:
\begin{equation}
    \mathrm{Pr}\left( \left| \hat{\theta} - \theta \right| \geq \epsilon \right) \leq 2 e^{-2 n \epsilon^2}.
    \label{eqn:basicHoeffdings}
\end{equation}
This inequality allows us to decide a value for this probability and then estimate the minimum number of labels needed to ensure that probability. 
Suppose we want the probability that we are off by more than $\epsilon$ to be no more than $\delta$. 
Then at least
\begin{equation}
    n \geq \frac{\ln (2/\delta)}{2 \epsilon^2}
    \label{eqn:numSamplesHoeffding}
\end{equation}
responses are needed to provide a bound on $\delta$.
(Note that tighter bounds than Hoeffding's may be used, but for simplicity here we focus on Eq.~\eqref{eqn:basicHoeffdings}; see the discussion for more.)

Our crowdsourcing goal for a given task is to determine if the unknown label $z$ is 1 or 0 (for now we suppress the dependence on task index $i$ and timestep $t$). 
The difference between our current estimate $\hat{\theta}$ and $1/2$ represents our weight of evidence towards this decision. If we are confident to some degree that our estimate $\hat{\theta}$ is different from $1/2$, then we are able to conclude the label of the task based on whether $\hat{\theta} > 1/2$ or $\hat{\theta} < 1/2$ and when we can draw that conclusion we can also deem the task complete.
Using Eq.~\eqref{eqn:numSamplesHoeffding} and our current estimate with $n$ responses,  we can then estimate how many additional responses $m$ we need until our confidence interval (or margin of error) does not include $1/2$:
\begin{equation}
    m \geq \frac{\ln(2/\delta)}{2\left(\frac{a}{n} - \frac{1}{2}\right)^2} - n.
    \label{eqn:n2estimate}
\end{equation}
Equation \eqref{eqn:n2estimate} shows us that the closer the task's parameter $\theta$ is to $1/2$, the more costly the task will be in terms of requiring more responses to distinguish if the label should be 0 or 1.
Of course, this estimate may be inaccurate as it relies on the current value of $\hat{\theta}=a/n$ at $n$ responses.
In reality, as more responses are gathered, $\hat{\theta}$ will be revised. 
These updated estimates can be automatically incorporated into this equation as new responses are received, yielding improved forecasts for $m$.

However, Eq.~\ref{eqn:n2estimate} is not valid when $\hat{\theta}=1/2$. In this scenario, we can ask: what if we receive our next response and it is $+1$ or it is $0$?
Since all we currently know in this scenario is $\hat{\theta}=1/2$, we should assume either outcome is equally likely, giving a revised estimate $\hat{\theta} = a/(n+1)$ (if the new response is 0) or $\hat{\theta} = (a+1)/(n+1)$ (if the new response is $+1$).
Thankfully,
$(\hat{\theta}-1/2)^2$
is the same in both cases, and so plugging either into Eq.~\eqref{eqn:n2estimate} will give the same estimate for $m$:
\begin{equation}
    m \geq \frac{\ln(2/\delta)}{2\left(\frac{a}{n+1} - \frac{1}{2}\right)^2} - n -1,
    \label{eqn:n2estimateAt1/2}
\end{equation}
where the $-1$ counts the additional label we assume we will receive.

In summary, we can estimate the number of additional responses $m$ needed to complete a task using
\begin{equation}
    m \geq \begin{cases} \frac{\ln(2/\delta)}{2\left(\frac{a}{n} - \frac{1}{2}\right)^2} - n &\mbox{if } a/n \neq 1/2, \\ 
\frac{\ln(2/\delta)}{2\left(\frac{a}{n+1} - \frac{1}{2}\right)^2} - n -1 & \mbox{if } a/n =1/2. \end{cases}
\label{eqn:n2overall}
\end{equation}

Once a task's $\hat{\theta}$ has been shown to be different statistically from $1/2$, the additional cost is $m\leq 0$ (no additional responses are needed). 
To use in subsequent sections, we define the set of \emph{available tasks} $M(t)$ as those where additional responses are needed: $M(t) = \{i : m_i(t) > 0\}$, where (suppressing the dependence on $i$ and $t$) $m_i(t)$ is given by Eq.~\eqref{eqn:n2overall}.

\subsection{Deciding when to request a new task}
\label{subsec:costforefasting:decidingnewtask}

The ability to estimate the cost to complete a task allows us to introduce a simple decision rule for when to request new tasks: \emph{request a new task when the expected cost to complete a new task is less than the estimated cost to complete the currently available task that is closest to completion}.

Specifically, let $i \in [1, \ldots, N_{t}]$ index the $N_{t}$ currently available tasks, and let $m_i$ be our current estimate for the cost to complete task $i$.
Let the expected cost to complete a new, unseen task be $\avg{n_j}$ (we compute this below).  
Comparing the $\{m_i\}$ with $\avg{n_j}$ then informs our decision rule for growing the set of tasks.

To decide whether or not to request a new task at some time $t$, we study two specific \emph{Growth Rules} (GRs):
Request a new task when
\begin{alignat}{2}
    \avg{n_j} &<  \min \{m_i\} \quad && \text{Growth Rule I (GR I)} \label{eqn:rg1}\\
    \avg{n_j} &< \mathrm{median} \{m_{i}\}     \quad && \text{Growth Rule II (GR II)}, \label{eqn:rg2}
\end{alignat}
where the minimum and the median are taken over the set of tasks for which additional responses are needed at time $t$, $M(t)$. 
We include the second rule (GR II) to provide a potentially less extreme counterpoint to GR I in that using the median as a decision point may be less influenced by outlier tasks than the minimum.

The intuition behind these growth rules is as follows.
As the crowd works on completing the currently available tasks, inexpensive tasks (those with $\theta$ far from 1/2) will finish first, and soon only expensive tasks (those with $\theta$ close to 1/2) will remain. Eventually, the remaining tasks will be costly enough that the crowdsourcer will be better off taking the chance on a brand new task.
Our experiments (Secs.~\ref{sec:methods} and \ref{sec:results}) investigate using these rules to elicit new tasks during crowd-generated microtask crowdsourcing.

\subsection{Estimating the cost to complete an unseen task}
\label{subsec:costforecasting:estimatingunseen}

Given the growth rules introduced in Eqs.~\eqref{eqn:rg1} and \eqref{eqn:rg2}, a question remains: how can we estimate the expected cost to complete a task $j$ when the task is unseen or has no responses (i.e., $a_j=n_j=0$)?
One option is to track the mean completion cost of previously completed tasks and use that for $\avg{n_j}$.
Another option is to track the mean \emph{parameter} $\hat{\theta}$ of previously completed tasks $\avg{\hat{\theta}}$ and use that mean within Eq.~\eqref{eqn:n2overall} to estimate the completion cost.
The former uses more data, but the latter option may be preferable as the GRs are then comparing two estimated costs instead of one observed cost and one estimated cost---if the estimates are biased then comparing two estimates may prevent or at least limit the bias from having a harmful impact.
However, here we take a simpler approach focused on computing the expected cost from only a given prior distribution of $\theta$.

Given a prior distribution $P(\theta)$ for task parameters, we can estimate the expected minimum cost to complete unseen tasks if they are sampled from that prior:
\begin{equation}
    E[n] \approx \int_{n_{\min}}^{\infty} n P(n) dn,
\end{equation}
where $n_{\min} \equiv 2\ln(2/\delta)$ is the expected minimum cost for the ideal case of $\theta=0$ or $\theta=1$. 
Here $P(n)$ can be derived by performing a change-of-variables on the prior distribution $P(\theta)$.

Unfortunately, $E[n]$ diverges for any $P(\theta)$ that assigns sufficient probability at or near $\theta=1/2$, as tasks at that $\theta$ will on average never be completed.
To ensure convergence, we assume a bound is used for the maximum amount of responses $n_{\max}$ that should be spent on a given task, and tasks $i$ that reach $n_i \geq n_{\max}$ without being deemed complete are abandoned. 
Although here we used this bound only theoretically (when computing $E[n]$) since Opt-KG itself helps to prevent over-spending~\cite{chen2013optimistic},
in practice this bound can prevent a growth in sunk costs where expensive tasks consume an inordinate amount of the crowdsourcer's budget.
We explore the effects of this bound below.

Using this bound, the expected minimum cost to complete unseen tasks can be estimated: 
\begin{align}
E[n] &= n_{\max} \eta \sqrt{2}  + 2\left(1-\eta\sqrt{2} \right)\int_{n_{\min}}^{n_{\max}}nP(n)dn
\label{eqn:expectedcostUnseenNoassume}\\
&= \sqrt{n_{\min}n_{\max}}\left(2-\eta \right)-n_{\min}\left(1-\eta\right),
\label{eqn:expectedcostunseen}
\end{align}
where $\eta \equiv \sqrt{n_{\min} /  n_{\max}}$ and the second line holds for a uniform (prior) distribution of $\theta$.

\medskip

Finally, Eq.~\eqref{eqn:expectedcostunseen} for $E\left[n\right]$ (or Eq.~\eqref{eqn:expectedcostUnseenNoassume} for a different prior) and Eq.~\eqref{eqn:n2overall} for additional costs $\{m_i\}$ can be used in our Growth Rules, Eqs. \eqref{eqn:rg1}--\eqref{eqn:rg2},
to perform cost forecasting for crowd-generated microtask crowdsourcing.

\section{Materials and Methods}
\label{sec:methods}

Here we describe the real and synthetic crowdsourcing datasets we apply cost forecasting to, how to perform crowd-generated crowdsourcing on these data, and we introduce a non-growth baseline control to understand the performance of cost forecasting.

\subsection{Datasets}
\label{subsec:datasets} 

We study three crowdsourcing datasets.
 These data were not generated using an efficient allocation algorithm, and so it has become standard practice to evaluate such algorithms with these data~\cite{li2016crowdsourcing,mcandrew2017reply}--- since labels were collected independently, one can use an allocation algorithm to choose what order to reveal labels from the full set of labels, essentially ``rerunning'' the crowdsourcing after the fact.
Due to generally small number of responses for each task in these datasets, to simulate a response from a worker to a task we sample from a Bernoulli distribution with a probability $\hat{\theta}$ that is estimated from the responses for that task given in the original data.

Below we describe each dataset and how to use these data with crowd-generated microtask crowdsourcing, where the set of tasks changes throughout the crowdsourcing.

\begin{description}
\item[RTE] Recognizing Textual Entailment~\cite{snow2008cheap}. 
Paired written statements from the PASCAL RTE-1 data challenge~\cite{dagan2006pascal}.
Workers were asked if one written statement entailed the other.
These data consist of $N=800$ tasks and $8,000$ responses, with each task receiving 10 responses.
Data are available at \url{https://sites.google.com/site/nlpannotations/}.

\item[Bluebirds] Identifying Bluebirds~\cite{welinder2010multidimensional}.
Each task is a photograph of either a Blue
Grosbeak or an Indigo Bunting, 
Workers were asked if the photograph contains an Indigo Bunting.
There are $N=108$ tasks and $4,212$ responses, with 39 responses for each task. 
Data are available at \url{https://github.com/welinder/cubam}.

\item[Games] This dataset contains crowdsourcing tasks generated from an app based on a TV game show, ``Who
Wants to Be a Millionaire''~\cite{CPE:CPE4168}. 
When a question is first revealed on the show, the app sends a task containing the question and 4 possible answers to the users. 
Responses from users and correct answers were collected.
Data were preprocessed and responses binarized following the procedure used by Li \emph{et al.} \cite{li2016crowdsourcing}. 
The dataset contains $N = 1,682$ tasks and $179,162$ responses.
Data are available at \url{https://github.com/bahadiri/Millionaire}.

\end{description}

To study crowd-generated microtask crowdsourcing on these datasets, we first sample $N_0$ tasks from the $N$ tasks in the dataset to construct the initial seed tasks for the crowdsourcer to use. 
To replicate requesting a new task, we simply draw from the set of tasks remaining in the dataset that have not yet been requested.
In other words, at the start of crowdsourcing there are $N_0$ tasks available to the crowdsourcer and $N-N_0$ tasks which are in the data but not yet requested. 
The growth rule in use determines when new tasks should be generated, simulating the crowdsourcer's decision process.
Crowdsourcing continues until the budget $B$ is exhausted or all $N$ tasks have been requested.
Budget is used to request new tasks and to receive responses to existing tasks.

\subsection{Synthetic crowdsourcing} \label{sec:methods:simulations}
We supplement our results from real crowdsourcing data by performing controlled simulations. 
We generate datasets following the model defined above by assuming each worker response to task $i$ follows a Bernoulli distribution with parameter $\theta_i$.
This controls for the cost of the task and the amount of responses needed to accurately label $\hat{z}_i=0$ or $\hat{z}_i=1$. 
This assumes workers are reliable; see the discussion for incorporating worker reliability.
Note also that $\theta_i$ is used only to simulate worker responses---all subsequent calculations are performed using the estimate $\hat{\theta}_i$ as $\theta_i$ itself is unknown to the crowdsourcer.
When tasks are created, we draw $\theta_i$ from a uniform prior distribution but we can also draw from other probability distributions such as the Beta distribution. 
To begin each run of crowdsourcing, we generate a set of $N_0$ seed tasks.
To simulate requesting a new task $j$ from a worker at time $t$, we draw a new $\theta_j$ from the underlying prior distribution, add $j$ to the set of tasks, increment the number of tasks $N(t+1) = N(t)+1$, and so forth.
Unless otherwise noted, in simulations, we used $N_0 = 100$ and a total budget (Sec.~\ref{subsec:background:efficientallocation}) of $B = 3000$; we explore the effects of these and other parameters in our experiments below.
Using this model, we can apply efficient budget allocation techniques such as Opt-KG and implement the growth rules defined above.

\paragraph{Baseline control}

To understand better the performance of cost forecasting,
for each Growth Rule, we compare to a \emph{non-growth baseline} that controls for the number of tasks and total budget spent on responses to those tasks. 
In this baseline, the number of tasks available at the start matches the final number of tasks generated when using cost forecasting, no new tasks are proposed by the crowd, and the budget available to the baseline is equal to the number of labeling responses received when using cost forecasting. 
Specifically, the budget for responses $B_r$ available to the baseline is $B_r = B-\left(N-N_0\right)$ where $B$ is the total budget used by cost forecasting and $N$ is the final number of tasks generated by the  crowdsourcing we are comparing against.
We perform one matching realization of the baseline for each realization of cost forecasting, as randomness in worker responses leads to variability in the total number of tasks proposed across different realizations of cost forecasting.
Note that this baseline is equivalent to a growth rule that performs all growth at the start of the crowdsourcing, then receives all worker responses to those tasks until the budget is exhausted. 
This contrasts with cost forecasting which dynamically alternates between growing tasks and responding to tasks using a given Growth Rule.

\section{Experiments}
\label{sec:results}

\subsection{Real and synthetic data}
\label{subsec:real-synthetic-data}

\begin{figure}[t]%
    \centering
    {\includegraphics[width=0.475\textwidth,trim=0 15 0 8,clip=true]{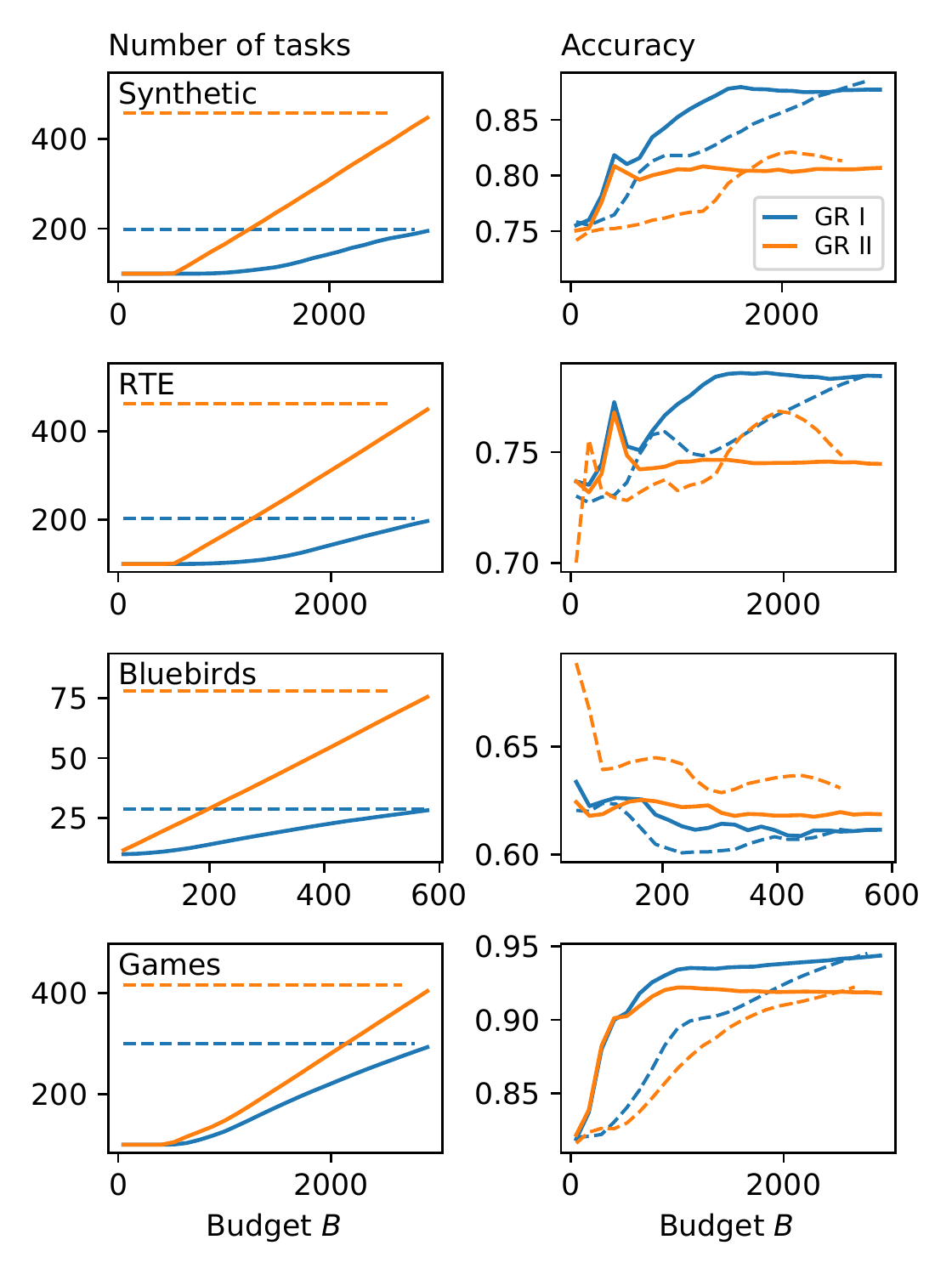}}
    \caption{Cost forecasting applied to synthetic and real world crowdsourcing data. 
    Accuracy of inferred labels is generally higher at given total budget for both growth rules (solid lines; blue: Growth Rule I, orange: Growth Rule II) than if all tasks were available to start (control, dashed lines).
    Higher accuracy at tight budgets allows cost forecasting to handle crowd-generated sets of tasks and to handle \emph{budget-uncertain scenarios} (see discussion), helping the crowdsourcer to ensure the gathered data is high-quality even if the budget is suddenly cut.
    \label{fig:big-fig-data}
    }
\end{figure}

We evaluate the performance of cost forecasting on simulated and real crowdsourcing data (Fig.~\ref{fig:big-fig-data}). 
Solid lines correspond to cost forecasting while dashed lines correspond to the non-growth baseline.
For these results we used cost forecasting parameters (Sec.~\ref{subsec:costforecasting:forecastingcost}) $\delta = 0.9$ for GR I, $\delta = 0.5$ for GR II (which exhibits faster growth than GR I), and  $n_{\max} = 10$ (Sec.~\ref{subsec:costforecasting:estimatingunseen}) for both; we further explore the dependence on $\delta$ and $n_{\max}$ below.
(Bluebirds, a smaller, noisier dataset, used $\delta = 0.5$ (GR I), $\delta = 0.1$ (GR II), $N_0=10$, $B=600$.)
Cost forecasting leads to slower growth at the beginning of crowdsourcing, visible in the long pause before the number of tasks begins to grow (Fig.~\ref{fig:big-fig-data}). 
Our method does not begin to grow until the crowd has provided enough responses about the seed tasks to achieve accurate labels.
In contrast, the non-growth baseline begins with all tasks initially available.
Examining the accuracy, or proportion of correct tasks, shows that cost forecasting achieves higher accuracy than the baseline for most data, especially for earlier in the budget, with Bluebirds (a difficult task with a global accuracy of only $\approx 0.65$) being a possible exception. 
Note that by controlling for the overall growth rate and budget of cost forecasting in the baseline (see above), the final accuracy (at high budgets) of both methods will on average always be the same, as both methods use the same Opt-KG allocation method.
Yet, cost forecasting can achieve higher accuracy at low budgets (often up to $\approx 5\%$) by dynamically determining the growth rate based on the past and current state of the crowdsourcing. 

\subsection{Dynamics of cost forecasting}

\begin{figure}[t]
    \centering
    {\includegraphics[width=0.475\textwidth,trim=0 15 0 3,clip=true]{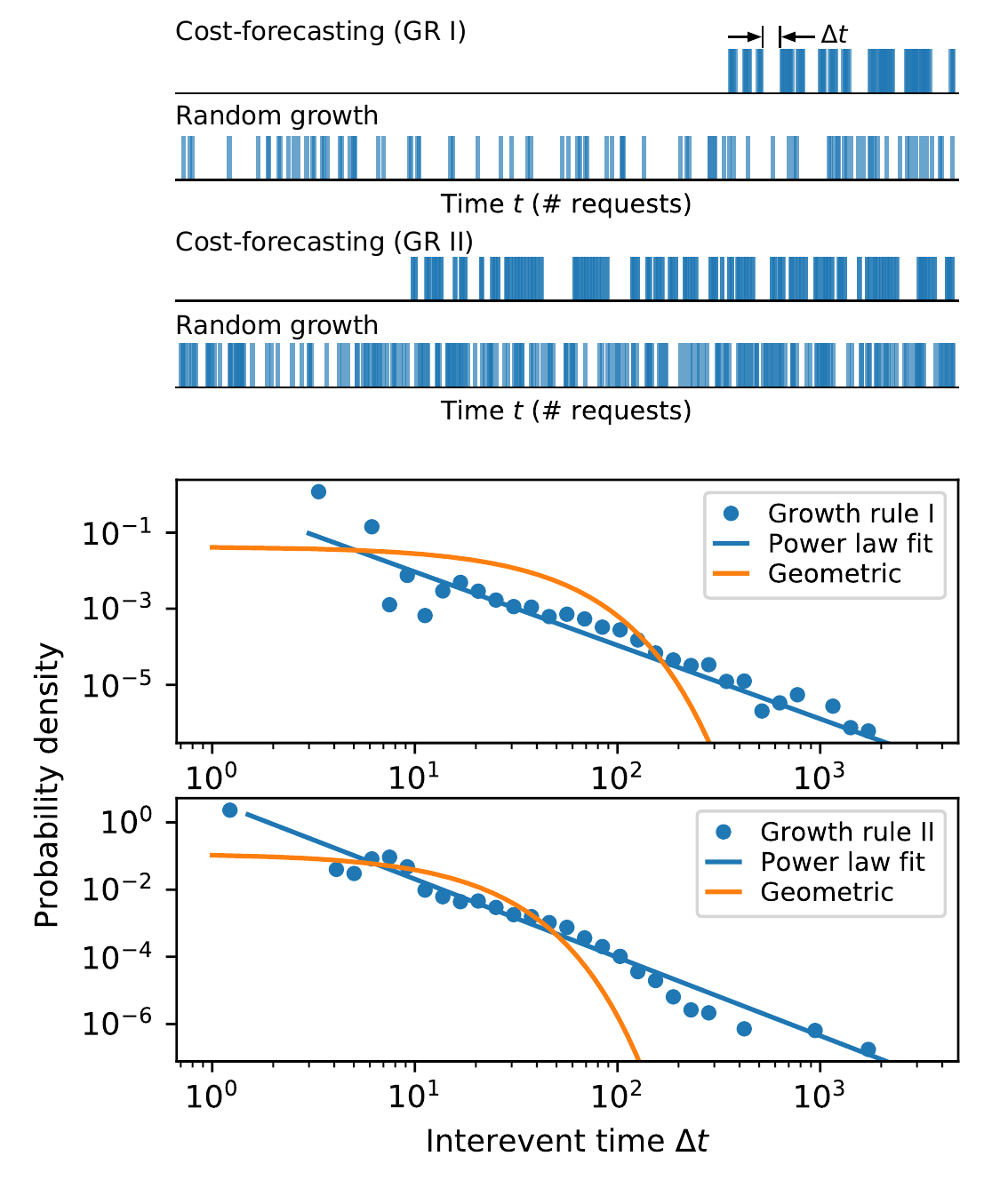}}
    \caption{Cost forecasting leads to a bursty pattern of growth.
    \lett{Top} 
    Example ``spike trains'' highlighting when new tasks are requested for one run of each growth rule. 
    For context, we show for each an example of a spike train with the same average growth rate where growth is equally likely to occur at any point.
    \lett{Bottom} 
    Cost forecasting leads to a heavy-tailed, approximately power-law distribution of $\Delta t$, the waiting times or interevent times between growth requests. 
    This distribution is characteristic of a bursty process, unlike the geometric distribution of $\Delta t$ displayed by a memoryless random growth process. 
    \label{fig:intervent-time}    
    }
\end{figure}

Cost forecasting decides between requesting responses to existing tasks and requesting new tasks. 
The dynamics of this decision process will vary as the responses are gathered for existing tasks, leading to a dynamical pattern distinctly different from that exhibited by, e.g., constant random growth (Fig.~\ref{fig:intervent-time}, top).

A well-established way to study these dynamics is through the \emph{interevent times} $\Delta t$, the number of non-growth requests that occur between growth requests.
If a discrete-time process is memoryless, where each request is equally likely to be a growth request, $\Delta t$ will follow a geometric distribution $P(\Delta t = k) = p (1-p)^{k}$ where $p$ is the probability for a growth event. 
This converges to an exponential distribution for a continuous-time process, $P(\Delta t) = \lambda e^{-\lambda \Delta t}$, with rate parameter $\lambda$.
In contrast, \emph{bursty processes} exhibit heavy-tailed, often power-law distributions of $\Delta t$: $P(\Delta t) \propto (\Delta t)^{-\alpha}$ for power-law exponent $\alpha > 1$~\cite{goh2008burstiness}.
Power-law distributions show higher probabilities relative to exponentials for both very short $\Delta t$ and very long $\Delta t$, capturing the long pauses of non-activity punctuated by sudden bursts of activity that are characteristic of bursty processes.

Figure \ref{fig:intervent-time} shows the interevent distribution for both cost forecasting growth rules.
At top, we use a ``spike train'' to illustrate the growth events around one run of simulated crowdsourcing, with another random growth spike train demonstrating a memoryless process where growth events occur at the same rate as the cost forecasting growth rule.
Below, we show power-law  and geometric distributions fitted to the $\Delta t$ observed over 50 runs~\cite{alstott2014powerlaw}.
Indeed, we see that cost forecasting is heavy-tailed and at least approximately well explained by a power-law distribution, indicating it is a bursty process.
Furthermore,
likelihood-ratio tests~\cite{alstott2014powerlaw} showed significant evidence ($p < 10^{-14}$) for power-laws over exponentials (the continuous analog of the geometric distribution) for both growth rules. %
The burstiness of cost forecasting shows that the algorithm tends to alternate between suddenly requesting multiple new tasks (short interevent times) and then focusing for some time on receiving responses to existing tasks (long interevent times).
In other words, it is reactive to the current state of the crowdsourcing, trading off expected costs given by responses to the current tasks with the potential cost a new, unseen task will require to be completed.

\subsection{Parameter dependence}
\label{subsec:parameter-dependence}

The cost forecasting procedure introduced in Eqs.~\eqref{eqn:n2estimate}--\eqref{eqn:expectedcostunseen} depends on parameters $\delta$ and $n_{\max}$.
Here we explore some effects of these parameters.
Further, we assume each crowd-generate microtask crowdsourcing begins with an initial seed of $N_0$ known tasks (and no responses), so we also study how cost forecasting behaves for different size seeds.

\begin{figure}[t]
    \centering
    {\includegraphics[width=0.47\textwidth,trim=0 11 0 0,clip=true]{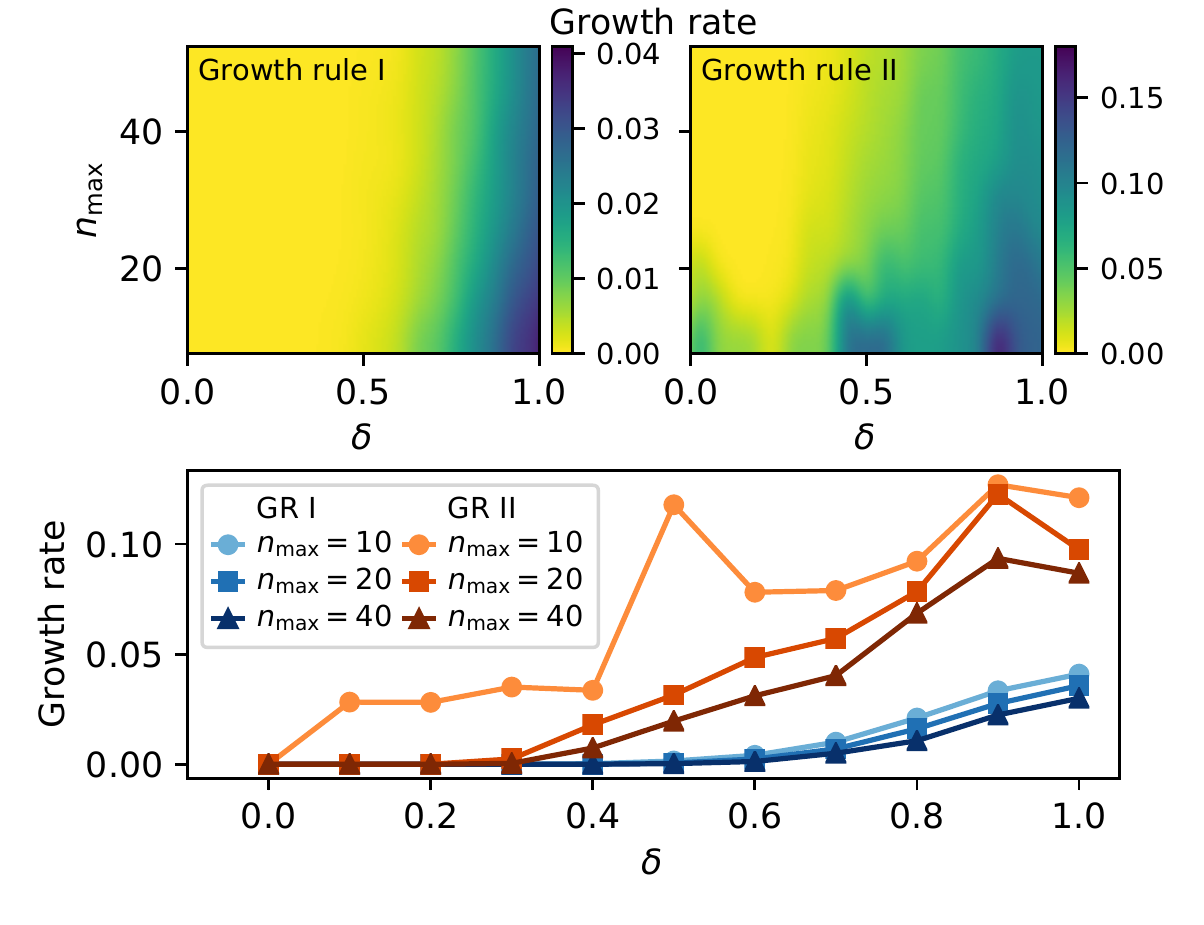}}
    \caption{How average growth rate depends on cost forecasting parameters $\delta$ and $n_{\max}$ (Sec.~\ref{sec:costforecasting}) for simulated crowdsourcing. 
    Generally, $\delta$ has a stronger effect than $n_{\max}$ on growth rate, especially GR I.
    \label{fig:growth-rate-parameter-dependence}}
\end{figure}

Figure \ref{fig:growth-rate-parameter-dependence} uses simulated crowdsourcing to explore the dependence of the average growth rate of tasks on $\delta$ and $n_{\max}$. 
Examining Fig.~\ref{fig:growth-rate-parameter-dependence}, $n_{\max}$ has little effect on GR I's growth rate while increasing $\delta$ provides the researcher with some ability to tune a given growth rule's growth rate.
In particular, using GR I and varying $\delta$ from 1/2 to 1 increases the typical growth rate by about 4\% (Fig.~\ref{fig:growth-rate-parameter-dependence}, bottom) essentially independently of $n_{\max}$. 
GR II, in contrast, exhibits a higher overall growth rate, a slightly greater dependence on $n_{\max}$ than GR I, and the growth rate increases by $\approx 8\%$ for $\delta = 1$ compared with $\delta = 0.1$ (Fig.~\ref{fig:growth-rate-parameter-dependence}, bottom).
These results show that the choice of $n_{\max}$
does not have a large impact on growth rate  for GR I, while GR II shows increased growth rate for small values of $n_{\max}$.

\begin{figure}[ht!]
    \centering
    {\includegraphics[width=0.475\textwidth,trim=0 13 0 10,clip=true]{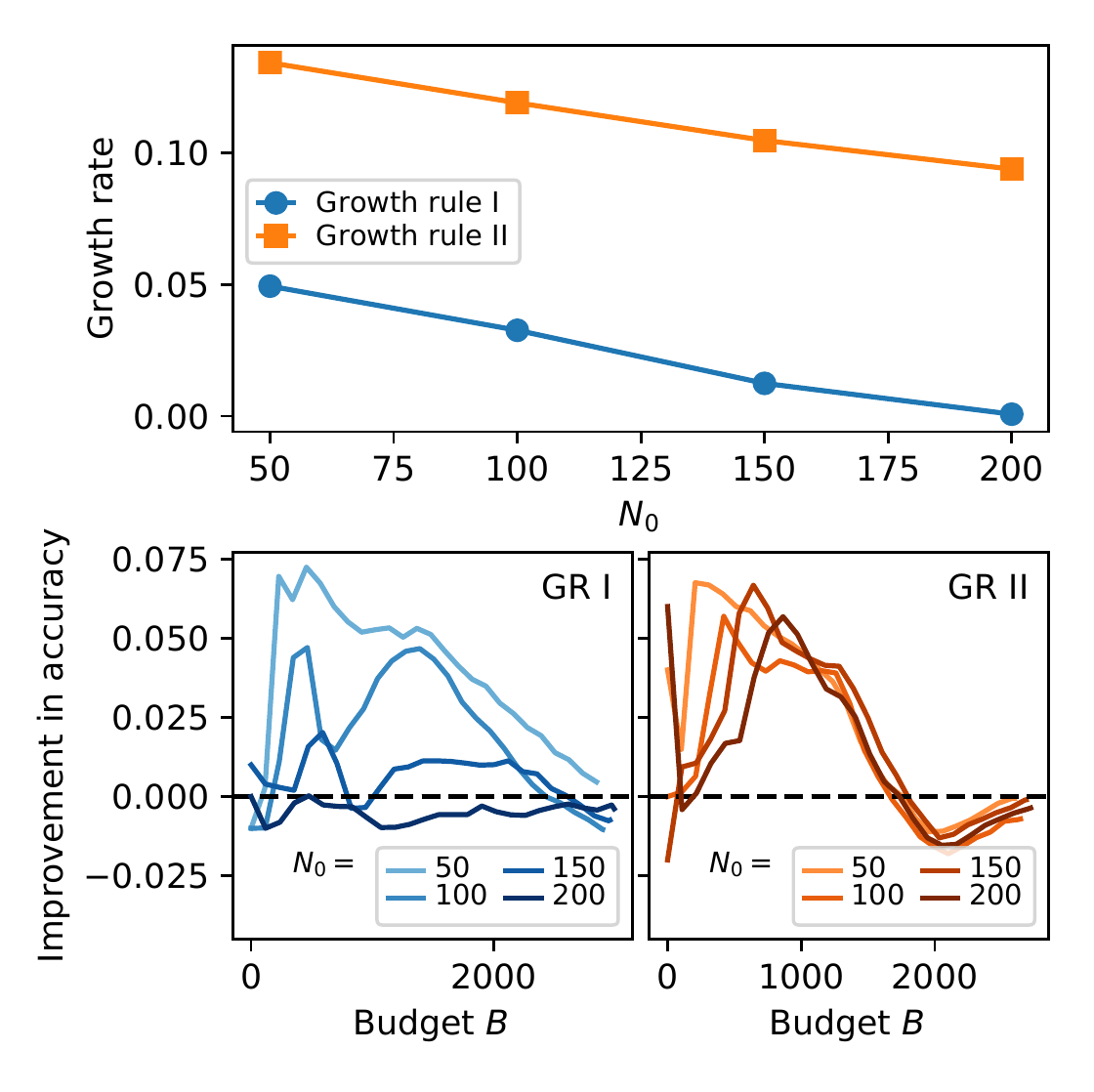}}
    \caption{Effect of initial number of tasks $N_0$ on growth rate and improvement in accuracy for simulated crowdsourcing.
    Generally, larger $N_0$ leads to less growth and less improvement in accuracy, since very large $N_0$ effectively acts like a fixed set of tasks.
    \label{fig:seed-size}
    } 
\end{figure}

We next investigate how growth rate depends on the initial number of available tasks $N_0$.
When many tasks are available to start, we anticipate that cost forecasting will spend more time exploring the available tasks before it begins to grow, which will lead to a lower overall growth rate for a fixed budget.
Indeed, Fig.~\ref{fig:seed-size} (top) shows that larger $N_0$ crowdsourcings have lower growth rates than smaller $N_0$ crowdsourcings for a given Growth Rule. 
For example, 
when $N_0 = 200$, the growth rate is approximately 5\% lower (for GR I) or 3\% lower (for GR II) than when $N_0 = 50$, indicating a small but potentially important affect on the overall crowdsourcing.

Given that larger $N_0$ gives lower growth rates, what effect does $N_0$ have on accuracy?
The bottom panels of Fig.~\ref{fig:seed-size} explore how accuracy improvement (accuracy of cost forecasting minus accuracy of corresponding baseline)
depends on different values of $N_0$. 
Generally, accuracy is improved at tight budgets using cost forecasting, but this improvement is lessened to some extent as $N_0$ increases---this is plausible as very large values of $N_0$ are effectively fixed-size traditional microtask crowdsourcings, meaning large $N_0$ are scenarios where there is less advantage for a crowdsourcer to apply cost forecasting.
Smaller $N_0$, however, show the advantages at tight budgets in terms of accuracy for cost forecasting.
We also note that (as in Fig.~\ref{fig:big-fig-data}) there is a consistent trend for GR II to briefly perform worse than the baseline at high values of $B$ ($\approx 2000$) before higher values of $B$ lead to comparable performance between the two approaches. 

\subsection{Non-stationary crowdsourcing---increasing completion costs}
\label{subsec:increasing-costs}

Our cost forecasting approach assumes the expected minimum cost to complete an unseen task is constant over the course of the crowdsourcing. 
Yet, is this a realistic assumption? 
One can imagine a scenario where the crowd initially proposes ``easy'' tasks (where consensus is reached quickly and the label can be inferred with few responses) then the crowd runs out of ``low-hanging fruit'' and later tasks will tend to be more expensive.
An example scenario is a question-answering site where all the easy-to-answer questions have already been proposed and subsequently proposed questions tend to be polarizing for the community.
If this occurs, how will it affect the performance of crowdsourcing using cost forecasting?

To explore how cost forecasting behaves under an increasing-cost scenario, we augment our crowdsourcing model by enabling the prior distribution for $\theta_i$, the probability of a 1-label for task $i$, to vary as more tasks are proposed by the crowd.
When this distribution becomes more sharply peaked at $\theta=1/2$, tasks will tend to be more costly to complete.
Then, to capture an increasing-cost scenario, we take a Beta distribution $\mathrm{B}(\alpha,\beta)$ for the prior of $\theta$ and make the parameters linearly increasing functions: $\alpha(N_t)=\beta(N_t) = 1 + s (N_t-N_0)$, where $N_t-N_0$ is the number of tasks proposed so far, $s$ parameterizes the rate at which tasks become more costly (as increasing $\alpha=\beta$ leads to a prior more sharply peaked at $\theta = 1/2$), and the intercept $1$ ensures the initial prior is a uniform distribution.

\begin{figure}
    \centering
    {\includegraphics[width=0.475\textwidth,trim=0 13 0 7,clip=true]{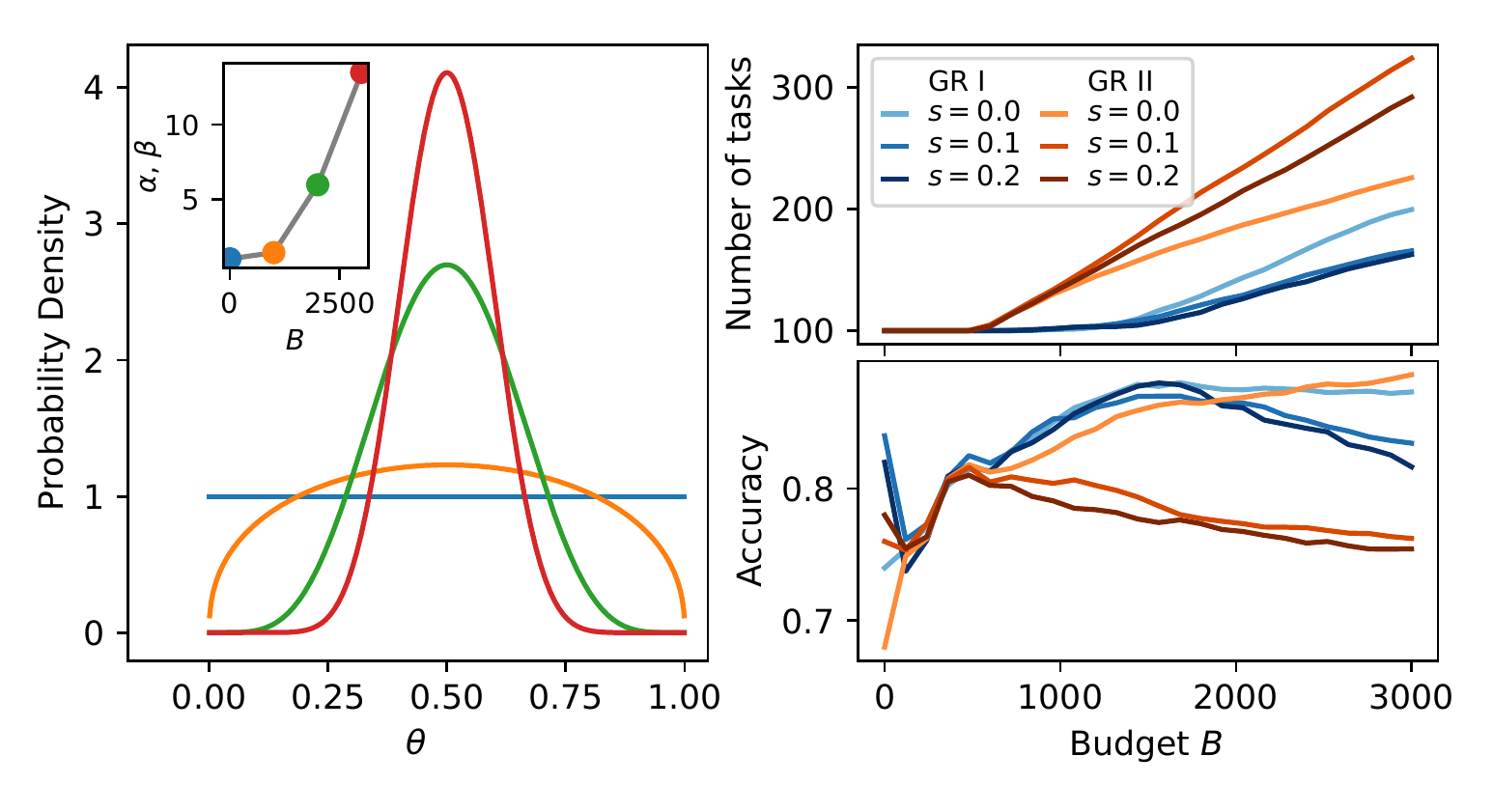}}
    \caption{Increasing completion costs.
    \lett{Left} The prior $P(\theta)$ for new tasks' 1-label probability $\theta$. 
    The cost to complete tasks grows as this distribution become more sharply peaked around $\theta = 1/2$ where it requires the most responses to distinguish 1- and 0-labels.
    \lett{Inset} The change in prior distribution parameters as crowdsourcing occurs. 
    The colored points correspond to the distributions shown in the main plot.
    \lett{Right}
    Accuracy for different rates of increasing cost $s$.
    Accuracy drops at high budgets for $s>0$, as expected, but both growth rules achieve similar accuracy for $s=0.2$ as they do for the less costly $s=0.1$.
    \label{fig:increasing-difficulty}}
\end{figure}

We illustrate the changing prior of the increasing-cost model in the left panel of Fig.~\ref{fig:increasing-difficulty}. 
In the inset of this panel we show how the Beta distribution parameters change as budget $B$ increases (and more new tasks are proposed), with the colored points in the inset corresponding to the distributions shown in the main plot.
In the right of Fig.~\ref{fig:increasing-difficulty} we illustrate how the growth rules perform as tasks of increasing cost are proposed---note that the cost forecasting method used here is not made aware of these changing costs.
Here we used $\delta = 0.5$ $(0.1)$ for GR I (GR II).
As we also saw in Fig.~\ref{fig:big-fig-data}, GRII generally exhibits more growth and lower accuracy than GRI, and we expect higher accuracy when there is lower growth as there will be more responses for fewer tasks.
This growth-accuracy tradeoff effect is exacerbated further here, when later tasks are more difficult than earlier tasks, as less growth leads to more responses to earlier, easier tasks. 
Indeed, accuracy drops at larger $B$ for higher $s$, as tasks become more difficult, but both growth rules handle the change in $s$ rather well, showing similar drops in accuracy for both $s=0.1$ and  the more costly $s=0.2$.
Yet GR II shows a faster growth rate for $s=0.1$ than $s=0.2$, demonstrating how, despite incorrectly assuming new tasks are always equally costly to complete, cost forecasting can still react to some extent to non-stationary task sets.

\section{Discussion}
\label{sec:discussion}

In this work, we introduced cost forecasting as a means to crowdsource crowd-generated microtasks where the crowd both completes tasks but also proposes new tasks to the crowdsourcer.
Crowdsourcing of crowd-generated microtasks can be used for question-answering sites, the design of new surveys, and in general can enable crowds to combine creative task proposal with traditional microtask work.
We demonstrated for binary labeling tasks on both synthetic and real-world crowdsourcing data that cost forecasting can leverage the performance of an efficient crowd allocation method and lead to improved accuracy.

Cost forecasting can also help budget-uncertain crowdsourcing. 
If a crowdsourcer does not know how many responses they will be able to gather, 
they will want to achieve and maintain a high accuracy as soon as possible, so that, whenever crowdsourcing terminates, the labels received for tasks are of as high a quality as possible.
One application of such budget-uncertain crowdsourcing is large-scale, automated A/B/n testing, where stopping rules may be evaluated online for many concurrent crowdsourcings.

There are many further directions to explore and extend this research.
One direction is the integration of cost forecasting with different crowd allocation methods.
We focused our validation on applying cost forecasting to Opt-KG, a popular and effective crowd allocation method for fixed sets of microtasks, free of parameters and focused on the overall accuracy of the generated task labels. 
Likewise, the statistical decision process of cost forecasting brings to mind Markov decision processes (MDP) and POMDP, and MDP and POMDP are common approaches to algorithmic crowdsourcing~\cite{DAI201352}.
Indeed, Opt-KG itself defines a policy using MDP~\cite{chen2013optimistic}; thus our results here demonstrate that cost forecasting can be fruitfully interfaced with MDPs.
More generally, as improved allocation methods are developed, it is important to examine if and how they can benefit from cost forecasting or other methods geared towards applying an allocation strategy to a set of crowd-generated microtasks. 
Developing methods that can directly allocate workers without assuming a fixed and known number of tasks would be an especially useful area of research.

Another direction for future research is to better understand how a crowdsourcer can integrate information about a particular crowdsourcing problem of interest.
For example, a crowdsourcer may already have a good idea about the difficulties of new tasks, perhaps from performing a pilot study. 
This information can be integrated into cost forecasting by choosing a non-uniform prior distribution for $\theta$. 
What about other cost forecasting parameters such as $\delta$, $n_{\max}$ or a different growth rule? 
A crowdsourcer will wish to balance their needs for accuracy and budget constraints when choosing these parameters.
Low-budget, pilot crowdsourcings may again be fruitful to help select these parameters and it is worth studying procedures for estimating their values.

Our formulation of cost forecasting is simple in several ways, but can be fruitfully extended.
We based our cost forecasting calculations on the Hoeffding bound for simplicity. 
This leaves considerable room for improvement as the Hoeffding bound is not particularly tight, and better results may be achieved using a tighter bound such as the \emph{empirical Bernstein inequality}~\cite{audibert2009exploration,maurer2009empirical}.
Further improvements include using a learning procedure where the estimated unseen task completion cost is dynamically learned as crowdsourcing is performed, although we found some support (Sec.~\ref{subsec:increasing-costs}) using an increasing-cost model that our basic cost forecasting procedure can already handle some changing costliness of new tasks.
We assume reliable workers, but worker reliability can be readily incorporating by using the worker reliability (or ``one-coin'') variant of Opt-KG or by incorporating worker reliability into whatever allocation method the crowdsourcer wishes to use.
We also assume the costs to request new tasks or request responses to existing tasks are the same, but of course in practice these may be different~\cite{sheng2008get}.
However, cost forecasting can automatically capture any task cost differential by modifying $E[n]$ to include a different proposal cost.
Likewise, the completion costs of unseen tasks are likely to vary over the course of a crowdsourcing, a phenomena we investigated using an increasing-cost model. 
While such models are useful, it is also important to understand how these costs may vary in practice (see \cite{Shtok:2012:LPA:2187836.2187939}). 
Do workers really run out of low-hanging fruit when performing crowd-generated microtask crowdsourcing? 
Experiments are needed to understand better how the set of tasks changes over time as the crowd proposes new tasks.

Finally, our cost forecasting Growth Rules focus on completion costs of tasks, as probabilistic cost estimators can be applied. 
Yet it would be especially interesting to use other quantities for growth rules. 
For example, if one can estimate the expected gain of novel information when requesting a new task, then a crowdsourcer can design crowd-generated microtask crowdsourcing to achieve goals such as crowdsourcing until a certain number of interesting or novel tasks are generated.

\section*{Acknowledgments}

{We thank Paul Hines and Hamid Ossareh for helpful comments.
This material is based upon work supported by the National Science Foundation under Grant No.\ IIS-1447634.}
~~~~

{\singlespacing

}

\end{document}